# Polychromatic forward-directed sub-Doppler emission from sodium vapour

Alexander M. Akulshin[1,2] *, Felipe Pedreros Bustos[1], Nafia Rahaman[2], and Dmitry Budker[1,3]

[1]*Johannes Gutenberg University, Helmholtz Institute, D-55128 Mainz, Germany*
[2] *Centre for Quantum and Optical Science, Swinburne University of Technology, PO Box 218, Melbourne 3122, Australia*
[3]*Department of Physics, University of California, Berkeley, CA 94720-7300, USA*
*Corresponding author: aakoulchine@swin.edu.au

The mechanisms responsible for ultraviolet to mid-infrared light generation in sodium vapours two-photon excited with continuous-wave sub-100 mW power resonant laser radiation are elucidated from orbital angular momentum transfer of the applied light to the generated fields. The measured 9.5 MHz-wide spectral linewidth of the light at 819.7 nm generated by four-wave mixing, sets an upper limit to the linewidth of two fields resulting from amplified spontaneous emission at 2338 nm and 9093 nm on population inverted transitions. Understanding details of this new-field generation is central to applications such as stand-off detection and directional laser guide stars.

## 1. INTRODUCTION

Coherent optical field generation in alkali vapours has a long history. One of the first lasers was realized in Cs vapours excited with a helium discharge tube [1]. Population inversion on two infrared transitions in Cs atoms making light amplification possible is the result of cascade spontaneous decay from higher levels. In the following years, pulsed lasers were used for more efficient pumping of alkali vapours and lasing effect was achieved in low-finesse cavities in most alkali atoms [2].

Later, upon strong optical excitation of dense alkali vapours, when $\Omega \gg \Delta_D$, where $\Omega$ and $\Delta_D$ are Rabi frequency and Doppler width, respectively, different kinds of nonlinear light-emitting mirrorless processes were observed and investigated [3]. For example, in dense sodium vapours, coherent infrared (IR) radiation in a vicinity of atomic and dimer transitions was obtained without optical cavities [4, 5, 6].

A variety of pump schemes, such as single-step, two-photon [4] or collision-assisted excitation [6] were considered. It was found that beside amplified spontaneous emission (ASE) [7, 8], processes of stimulated Raman scattering [9], degenerate and nondegenerate four-wave mixing (FWM) [3, 10] can co-exist and compete [11].

Nonlinear light-emission processes were the subject of a considerable amount of research motivated not only by possible applications, such as developing sources of tuneable infrared radiation, but also by interest in better understanding of the key processes responsible for new-field generation. In particular, Dicke superradiant emission in optics was achieved on several cascading IR transitions in sodium vapours [12]. The idea of two-photon laser was realized in strongly excited lithium vapours [13].

Emergence of highly coherent tuneable diode lasers made possible to precisely address particular optical transitions. In the case of resonant excitation using low-power continuous-wave (cw) lasers ($\Omega < \Delta_D$), optical coherences in multilevel atomic media driven by laser fields and quantum interference between different excitation pathways might significantly modify the light-emitting processes [14, 15]. Of particular interest is the case of quantum gases where intriguing new features of nonlinear mixing processes are predicted [16].

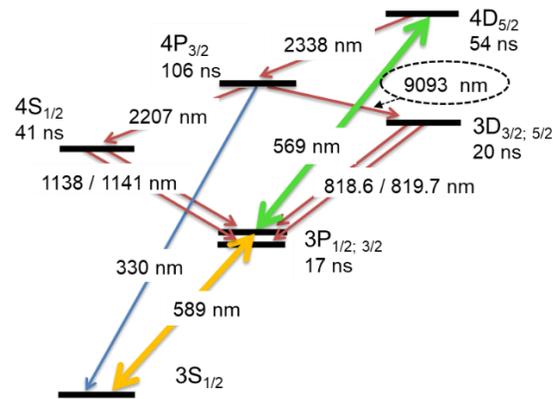

Fig. 1. Na atomic energy levels involved in the new-field generation in a sodium vapour two-photon excited to the $4D_{5/2}$ level by resonant laser light at 589 and 569 nm.

Potential applications in quantum information processing [17] revived general interest in frequency up- and down-conversion of temporally and spatially coherent radiation from one spectral region to another [18, 19, 20, 21]. Also, some promising applications such as standoff detection [22] and directional laser guide stars [23, 24] require minimization of power and atom-number thresholds for mirrorless lasing. This can be done with a model that includes important details of the interplay of parametric and non-parametric nonlinear processes.

Here we present the results of our study of polychromatic forward-directed emission of sodium vapour excited with cw laser light in the sub-100 mW power range, paying particular attention to identifying major processes involved in its generation. Investigating the spectral and spatial characteristics of the directional backward emission at 2206 nm in Na atoms excited to the $4D_{5/2}$ level by co-propagating laser light at 589 and 569 nm [25], we found that the forward-directed emission is polychromatic, as in the case of dense atomic vapours and strong pulse excitation [3-5].

## 2. EXPERIMENTAL SETUP

The experimental setup for studying forward-directed cw polychromatic light generation in sodium vapours is essentially the same as that described in our previous paper [24]. A simplified optical scheme of the experiment is shown in Figure 2.

A 10-cm long quartz cell containing metal sodium without buffer gas is used in the experiment. A resistive heater provides a temperature gradient along the cell preventing sodium atoms from condensing on the cell windows. The Na atom number density is estimated based on the measured temperature in the coldest part of the cell [26].

Two cw single-mode dye lasers (Coherent 699-21 and 899-21 lasing at 589 and 569 nm, respectively) are used for excitation of Na atoms to the $4D_{5/2}$ level. The upper limit for the short-term laser linewidth of both the lasers estimated using a confocal Fabry-Perot (FP) cavity is approximately 5 MHz.

The light from the two lasers is combined on a non-polarizing beam splitter to form a bichromatic laser beam that was loosely focused inside the cell. The minimum diameter of the gain region is approximately 250 µm. Photodiodes and photomultiplier tubes (PMT) are used for IR and ultraviolet (UV) light detection, respectively. To enhance the signal-to-noise ratio of observed signals, mechanical chopping of the 569 nm beam and lock-in detection are implemented when necessary.

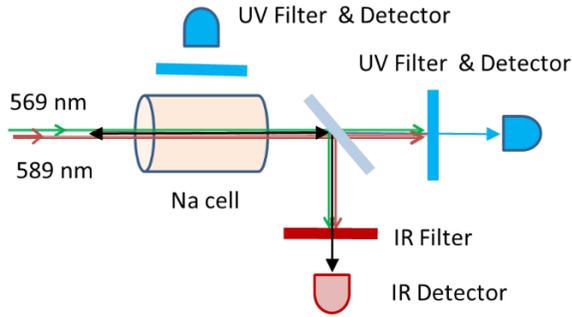

Fig. 2. Simplified optical scheme of the experiment.

For tuning the 569 nm laser to the two-photon resonance, UV or near-IR fluorescence emitted in the decay cascade of the Na atoms is monitored. The polarization and intensity of the 589 nm and 569 nm components are controlled with wave plates and polarizers. The maximum power of the 589 and 569 nm components of the bichromatic beam just before entering the cell is 60 and 40 mW, respectively.

As laser light reflected from the cell back window is intense enough for driving two-photon excitation in the counter-propagating geometry, this might significantly modify the spectral and spatial characteristics of the generated fields [27]. To avoid such complications, the Na cell is tilted by ~10 degrees reducing the overlap of the reflected and incident beams.

Ground-state re-pumping that is essential for more efficient excitation and ASE threshold reduction [28] is produced by a spectral sideband generated by frequency modulation of the 589 nm laser light at 1.713 GHz.

## 3. RESULTS AND DISCUSSION

### A. Directional UV emission

Fluorescence is a common indicator of the number of excited atoms. Analysing emission at 330 nm radiated in the longitudinal and transverse directions by sodium atoms two-photon excited to the $4D_{5/2}$ level, we find that at certain experimental parameters, the UV intensity in the co-propagating direction is significantly stronger than the backward or transverse emission. Figure 3 shows the power of UV emission in the transverse and forward directions as a function of frequency detuning of the 589 nm component of the applied laser light from the $3S_{1/2}(F=2)$-$3P_{3/2}(F'=3)$ transition. The side-peaks are due to excitation produced by sidebands of the frequency modulated laser light at 589 nm. Comparing the spectral profiles of UV emission in the forward and transverse directionswe, we find that for large red or blue detuning ($|\Delta\nu|$ >1 GHz) the spectral dependences of UV emission are similar, while at resonance, ($|\Delta\nu|$ < 0.5 GHz), a sharp enhancement occurs only for the co-propagating emission.

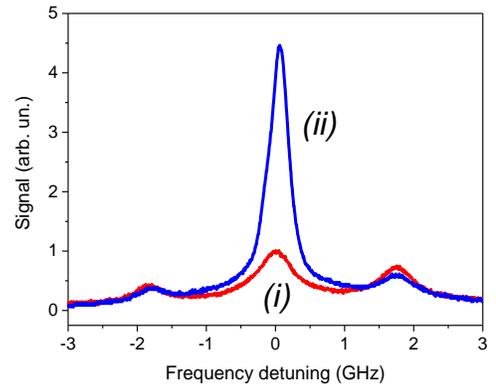

Fig. 3. Spectral dependences of UV emission in (*i*) transverse and (*ii*) co-propagating directions. UV power is shown as a function of frequency detuning of the 589 nm component from the $3S_{1/2}(F=2)$-$3P_{3/2}(F'=3)$ transition. The frequency of the 569 nm laser is tuned to the $3P_{3/2}$-$4D_{5/2}$ transition. The applied laser power at 589 and 569 nm is 20 and 12 mW, respectively. Atom number density $N$ in the cell is approximately $5.5\times10^{11}$ cm$^{-3}$.

The directional emission on the $4D_{5/2}$-$4P_{3/2}$ transition can be generated by the ASE process. Even though the natural lifetime of the $4P_{3/2}$ level is longer compared to the $4D_{5/2}$ level lifetime, cw population inversion on the $4D_{5/2}$-$4P_{3/2}$ transition is possible. This occurs because of the favorable branching ratio of the $4D_{5/2}$ level: spontaneous depopulation of the $4P_{3/2}$ level occurs faster than the decay from the $4D_{5/2}$ level to the $4P_{3/2}$ level [29]. As a result, population-inverted atoms contained in a significantly elongated region, determined by the length of the cell and the laser beam cross section in Na vapour, can generate directional emission on the $4D_{5/2}$-$4P_{3/2}$ transition at 2338 nm. To discriminate the forward-directed emission at 2338 and 2207 nm, we use an interference filter with 73 nm wide transmission window centered at 2350 nm. We find that the power and atom density threshold values at 2338 nm are even lower than for the ASE at 2207 nm.

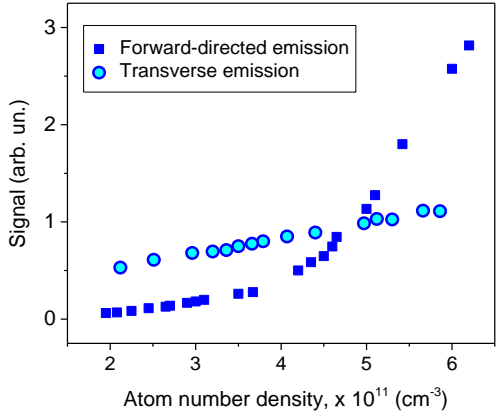

Fig. 4. Forward-directed and transverse emission at 330 nm as a function of atom number density $N$ in the cell. The frequency of the 589 nm laser is tuned to the $3S_{1/2}(F=2)$-$3P_{3/2}(F'=3)$ transition, while the 569 nm laser is tuned to the $3P_{3/2}(F'=3)$-$4D_{5/2}$ transition. The applied laser power at 589 and 569 nm is 30 mW and 12 mW, respectively. The transverse emission curve is offset vertically for clarity.

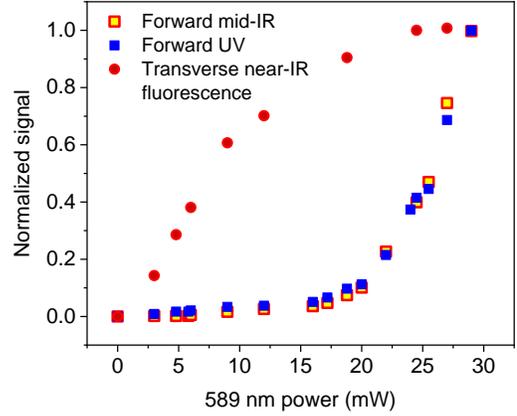

Fig. 5. Forward-directed emission at 330 nm and 2338 nm as well as combined near-IR fluorescence transmitted through the 820 nm interference filter as a function of the applied laser power at 589 nm, while the applied laser power at 569 nm is 15 mW. The frequency of the 589 nm laser is tuned to the $3S_{1/2}(F=2)$-$3P_{3/2}(F'=3)$ transition, while the 569 nm laser is tuned to the $3P_{3/2}(F'=3)$-$4D_{5/2}$ transition. The atom number density in the cell is $5\times10^{11}$ cm$^{-3}$.

The forward-directed UV radiation has a pronounced threshold-type dependence on the atom number density $N$ in the cell, as Figure 4 demonstrates. The enhanced co-propagating emission at 330 nm is directional. Spatial properties are evaluated by scanning the detector with an appropriate diaphragm across the beam. The typical divergence half-angle measured at the $1/e^2$ intensity level of the Gaussian fit to experimental profiles is approximately 8 mrad.

All these features of the forward-directed emission at 330 nm suggest that the UV radiation is the product of parametric FWM that involves the applied laser light and the directional light at 2338 nm generated on the $4D_{5/2}$-$4P_{3/2}$ transition in the co-propagating direction.

Figure 5 shows that normalized power dependences of forward-directed emission at 330 nm and 2338 nm are remarkably close, supporting the FWM mechanism of directional UV light generation. The combined near-IR fluorescence at 818.6 and 819.7 nm detected simultaneously in the transverse direction shows that the number of excited atoms does not increase sharply in the 15-25 mW power range.

We note that this frequency up- and down-conversion in Na vapours is an analogue of the mid-IR and blue light generation with Rb vapours [14].

## B. Multiple wave-mixing loops

In addition to the $4D_{5/2}$-$4P_{3/2}$ and $4P_{3/2}$-$4S_{1/2}$ transitions, there are other possibilities for creating population inversion and wave mixing in sodium vapours excited to the $4D_{5/2}$ energy level. The co-propagating light at 2338 and 2207 nm together with the applied laser field at 569 nm connect the $3P_{3/2}$, $4D_{5/2}$, $4P_{3/2}$ and $4S_{1/2}$ levels forming one more parametric FWM loop. In this case, if phase matching is achieved, co-propagating emission at 1141 nm on the $4S_{1/2}$-$3P_{3/2}$ transition may be generated. Also, considering the natural lifetime of the $4S_{1/2}$ and $3P_{1/2}$ levels (Fig. 1), steady-state population inversion should occur on the $4S_{1/2}$ - $3P_{1/2}$ transition resulting in possible ASE at 1138 nm.

Atoms from the $4P_{3/2}$ level spontaneously decay to both the $3D_{3/2}$ and $3D_{5/2}$ levels with low combined probability (< 2 %), while the decay to the $3S_{1/2}$ and $4S_{1/2}$ levels is much more probable. However, as the lifetimes of the $4P_{3/2}$ and $3D_{5/2}$ levels are favorable for creating a large steady-state population inversion on the $4P_{3/2}$-$3D_{3/2}$ and $4P_{3/2}$-$3D_{5/2}$ transitions, directional ASE at 9093 nm may be generated.

The currently used Na cell does not allow detecting the radiation at 9093 nm because the quartz windows are opaque in this spectral region. However, the presence of ASE on the $4P_{3/2}$-$3D$ transition is revealed indirectly via one more FWM process. Indeed, the applied laser field at 569 nm and the co-propagating ASE fields at 2338 and 9093 nm can form two zig-zag mixing loops ($3P_{3/2}$-$4D_{5/2}$-$4P_{3/2}$-$3D_{3/2}$-$3P_{3/2}$ and $3P_{3/2}$-$4D_{5/2}$-$4P_{3/2}$-$3D_{5/2}$-$3P_{3/2}$) that may produce directional light on the $3D_{3/2}$-$3P_{3/2}$ and $3D_{5/2}$-$3P_{3/2}$ transitions at 819.704 and 819.7077 nm, respectively, according to [30]. We note that population inversion on the $3D_{3/2}$-$3P_{1/2}$ transition also may generate directional ASE at 818.550 nm.

Figure 6 demonstrates that at certain experimental conditions two forward-directed near-IR fields are simultaneously generated. They transmit through interference filters with 10 nm wide transmission windows centered at 1140 or 820 nm, respectively. Both new optical fields show pronounced threshold-type dependence on the applied laser power at 569 and 589 nm, as Figure 7 demonstrates. The directional light in the spectral range around 1140 nm is saturated just at the 569 nm threshold power of the near-IR emission around 820 nm, suggesting a link between two mixing loops via their phase matching conditions. The power of forward-directed radiation that passes through the 820 nm interference filter is 30 μW, when the laser power at 589 nm and 569 nm is 40 mW and 30 mW, respectively, and the atom number density in the cell is approximately $8\times10^{11}$ cm$^{-3}$.

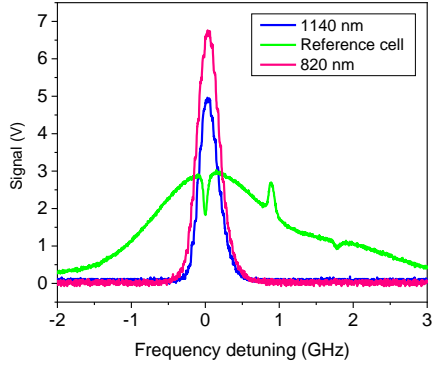

Fig. 6. Forward-directed emission in the spectral range of 1140 nm and 820 nm, as well as the reference saturated fluorescence at 589 nm detected in an additional cell, as a function of frequency detuning of the laser light at 589 nm from the $3S_{1/2}(F=2)$-$3P_{3/2}(F'=3)$ transition, while the 569 nm laser is tuned to the $3P_{3/2}(F'=3)$-$4D_{5/2}$ transition. The applied laser powers at 589 and 569 nm is 40 mW and 25 mW, respectively. The atom number density in the cell is $7\times10^{11}$ cm$^{-3}$.

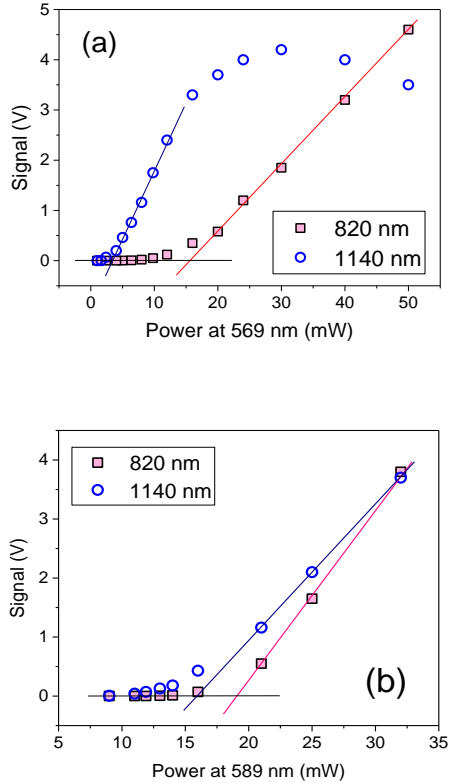

Fig. 7. Power dependences of forward-directed emission transmitted through the interference filters with 10 nm wide transmission bands centered 820 nm and 1140 nm as a function of the applied laser power at (a) 569 nm and (b) 589 nm, while the laser powers at 589 nm and 569 nm are 40 mW and 30 mW, respectively. The 589 nm laser is tuned to the $3S_{1/2}(F=2)$-$3P_{3/2}(F'=3)$ transition, while the 569 nm laser is tuned to the $3P_{3/2}(F'=3)$-$4D_{5/2}$ transition. The atom number density in the cell is $7\times10^{11}$ cm$^{-3}$.

## C. Orbital angular momentum transfer

Threshold-type power characteristics are common to nonlinear processes, but this fact does not help to discriminate which process, ASE or parametric FWM, is responsible for new-field generation. For distinguishing their origin, we conducted an experiment using vortex light [31] as was suggested in [32].

Every photon of vortex light carries $|\ell|\hbar$ orbital angular momentum (OAM), where $\ell$ is an integer, also known as optical topological charge. As OAM is conserved in the process of FWM [33], we check the OAM transfer from vortex laser light to the generated fields.

As was found earlier [34, 35], the frequency up-converted field, namely, collimated blue light generated by parametric FWM in Rb vapours, accumulates the total OAM of the applied laser light. We explain this, assuming that the ASE and FWM processes occur sequentially, by different involvement of laser photons from the applied vortex light in processes responsible for new-field generation. In the case of the ASE that produces the down-converted field at 5.23 μm, laser photons participate in the process indirectly, via creating population inversion on the appropriate transition. Thus, the OAM carried by the applied twisted laser light is transferred to atoms rather than to the generated field. Contrary, in the parametric FWM process, photons of the newly generated mid-IR field are directly involved in mixing with laser photons, and the resulting field accumulates the total OAM of the applied laser light, conserving the orbital angular momentum.

We note that a possibility of sharing OAM between frequency up- and down-converted fields and, consequently, OAM-entangled photon pairs generation, was discussed recently [36]. This was suggested based on an assumption that both fields are generated simultaneously and that the observed asymmetry in the OAM distribution originates from the different spatial overlap of the applied and generated fields. The role of spatial overlapping of optical fields involved in parametric generation in nonlinear media was theoretically studied in the seminal work [37]. This means that OAM can, in principle, be transferred to the frequency down-converted field. However, analytical consideration and quantitative experimental evaluations using the developed method for mode structure analysis of Laguerre-Gaussian beams generated by FWM in Rb vapours shows that for small OAM (<3) vortices transfers only to the frequency up-converted field [36]. Thus, even in the case of strong coupling of the ASE and FWM processes, we can conclude that the OAM transfer from the applied vortex light at 569 nm to the radiation passing through the 820 nm interference filter unambiguously identifies its generation mechanism.

Figure 8 shows transverse profiles of the emission transmitted through in the 820 nm interference filter in two different cases. The cross section of the beam produced with plane-wavefront laser light has a Gaussian-type profile (Fig. 8a). If a single-charge phase mask is introduced into the bichromatic beam before entering the Na cell, the spatial profile of the near-IR emission becomes doughnut-shaped (Fig. 8b). However, this shape is a necessary but insufficient condition for the identification of a vortex beam. To check if the 820 nm emission does indeed carry OAM and does not just reproduce the pipe-type shape of the gain region, we applied a simple method for OAM evaluation that is based on astigmatic transformation of an optical vortex beam [38, 39]. The number of high-contrast dark stripes across the vortex beam image produced by a tilted lens in the vicinity of the focal plane is equal to the topological charge $\ell$, or units of OAM, carried by light, while positive and negative inclination corresponds to clockwise and counter-clockwise rotation of the wave vector. Thus, a single high-contrast inclined dark band across the image (Fig. 8c) indicates the radiation under test is a single-charged vortex light.

By placing the single-charge phase mask solely into the 589 or 569 nm beam before combining on the beam splitter, we find that OAM can be transferred to the near-IR radiation only from the 569 nm laser pump component, as shown in Figure 9. Indeed, the tilted lens image of the 820 nm light (Fig. 9a) reveals a single dark stripe with the same inclination as in the 569 nm beam (Fig. 9c). Contrary, the smooth image (Fig. 9b) that does not show any brightness reduction in the centre unambiguously suggests that the beam is not vortex-bearing. This means that the OAM has not been transferred from the laser field at 589 nm. This pump field serves only for exciting Na atoms to the $3P_{3/2}$ level and its photons are not directly involved in the coherent wave-mixing process.

Thus, the observed near-IR directional radiation is the product of the FWM process in the $3P_{3/2}$-$4D_{5/2}$-$4P_{3/2}$-$3D_{5/2}$-$3P_{3/2}$ zig-zag mixing loop and must be at 819.7077 nm. This conclusion is supported by direct wavelength measurements performed using a High Finesse WS6-600 wavemeter. The measured wavelength of the generated light coincides with the table value for the $3P_{3/2}$-$3D_{5/2}$ transition [30]. This measurement also indicates that the forward-directed ASE is generated on the $4P_{3/2}$-$3D_{5/2}$ transition at 9093 nm.

The same procedure for origin identification is applied to the emission that passes through the 1140 nm interference filter. We find that it is the product of FWM that involves the laser light at 569 nm and two ASE fields at 2338 and 2207 nm in the $3P_{3/2}$-$4D_{5/2}$-$4P_{3/2}$-$4S_{1/2}$-$3P_{3/2}$ mixing loop. The emission at 1141 nm generated in Na vapours is an analogue of the 1366 nm directional emission observed in Rb vapours [27].

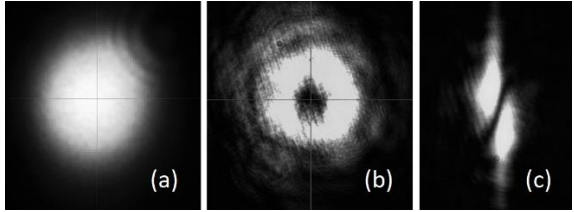

Fig. 8. Profiles of forward-directed near-IR emission generated (a) with laser light with plane wavefront and (b) vortex laser light. (c) Tilted-lens images of the near-IR light produced when both laser components at 569 and 589 nm are twisted. The 589 nm laser is tuned to the $3S_{1/2}(F=2)$-$3P_{3/2}(F'=3)$ transition, while the 569 nm laser is tuned to the $3P_{3/2}(F'=3)$-$4D_{5/2}$ transition. The atom number density in the cell is approximately $7 \times 10^{11}$ transition.

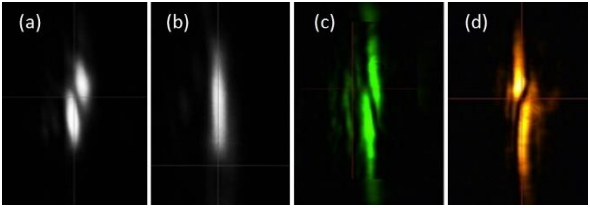

Fig. 9. (a, b) Tilted-lens images of the 820 nm light generated with the applied vortex laser light solely at 569 and 589 nm, respectively, while (c and d) profiles reveal topological charges of the applied vortex light at 569 and 589 nm, respectively. The opposite sign of OAM of the 589 nm beam is due to the phase plate orientation.

## D. Temporal coherence

Understanding spectral properties of directional emission is important for many potential applications. Currently we are unable to analyse the spectral width of the mid-IR fields directly, however, an upper limit for their linewidth for some wavelengths can be estimated from spectra of the 819.7 nm field. Both the ASE fields at 2338 and 9093 nm are involved in the parametric FWM process contributing to the spectral width of the 819.7 nm light.

For assessing the linewidth of the 819.7 nm light, we use a tuneable Fabry-Perot (FP) cavity. In this spectral range, the spectral resolution of the cavity evaluated using a highly coherent diode laser (Toptica DL Pro) is approximately 4.5 MHz.

Figure 10 shows the FP cavity transmission profiles of two 819.7 nm fields generated simultaneously in spatially separated regions inside the cell by two bichromatic laser beams from the same lasers, as shown in the figure insert. The profiles are recorded consecutively within an approximately 10 s time interval. The profiles demonstrate a similar shape that consists of three partially overlapping peaks. We attribute this structure to the hyperfine structure of the levels involved in the interaction.

The profiles show that the optical frequencies at the peak of each FWM field are separated by approximately 10 MHz, as the FP cavity low-frequency drift (approximately 0.3 MHz/min) is much smaller within the 10 s time. We explain this frequency shift by slightly different phase matching conditions in the interaction regions. Indeed, the bichromatic beams prepared from the same lasers have similar but not identical power ratio between the 589 and 569 nm components inside the cell resulting in different nonlinear refractive indexes of sodium vapours seen at different wavelengths. This makes the phase matching condition satisfied at slightly different optical frequency of the generated fields. Also, the independently generated mid-IR radiation at 2338 and 9093 nm with slightly different optical frequencies contributing to the observed frequency difference of the FWM fields.

The narrowest structure in curve (i) (Fig. 10) can be fitted with a Lorentzian function with 9.5 MHz width. As active synchronization of the two independent lasers at 589 and 569 nm is not applied; from the presented observations we can conclude that the upper limit for the sum linewidths of the IR fields at 9093 and 2338 nm is under 10 MHz, assuming that the lineshape of all fields is Lorentzian.

Ultimate temporal coherence of the optical fields produced by the processes of ASE and FWM as well as their atom-density and laser-power characteristics will be examined in detail using optical heterodyning in our future work.

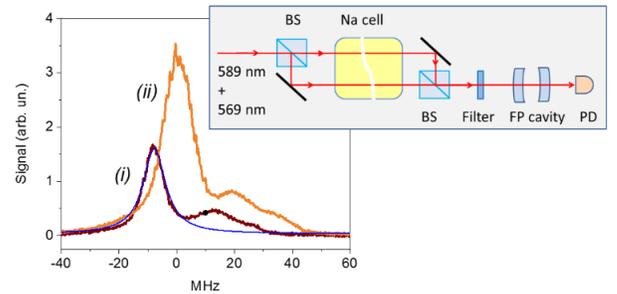

Fig. 10. Tunable Fabry-Perot cavity transmission profiles of two optical fields generated at 819.7 nm by FWM and recorded with frequencies of both the lasers fixed and tuned to the $3S_{1/2}(F=2)$-$3P_{3/2}(F'=3)$ and $3P_{3/2}(F'=3)$-$4D_{5/2}$ transitions, respectively. The atom number density in the cell was $6 \times 10^{11}$ cm$^{-3}$. Blue profile is a 9.5 MHz-wide Lorentzian fit to the largest transmission peak in curve (i).

## 4. CONCLUSION

In this work, we have shown that Na vapours two-photon excited to the $4D_{5/2}$ level with co-propagating laser light at 589 and 569 nm emit forward-directed polychromatic coherent radiation ranging from UV to mid-IR spectral regions. Directional ASE at 9093 nm is detected indirectly via the parametric FWM process resulting in generation of directional light at 819.7 nm.

All optical fields produced by ASE and FWM processes reveal pronounced threshold-type dependences on the atom number density and applied laser power.

Orbital angular momentum conservation is used to distinguish nonlinear processes responsible for new field generation in Na vapour.

The recorded spectral width of the new field at 819.7 nm generated by FWM puts a 10 MHz upper limit on the linewidth of all optical fields involved in the nonlinear mixing process, namely, the 589 and 569 nm laser radiation as well as the ones generated in the Na cell at 2338 and 9093 nm.

Very recently, the generation of UV light at a wavelength of 311 nm upon excitation of Rb atoms to a low-lying Rydberg state was reported in [40]. Pushing the limits of frequency up- and down-conversion to hard UV (below 280 nm) and to a level of several THz, respectively, using the two-photon excitation in alkali atoms, can be of great importance for a number of applications.

**Funding Information.** This work is partially supported by the US Office of Naval Research, Global (N62909-16-1-2113).

**Acknowledgment**. We thank Russell Mclean for useful discussions. F. P. B. acknowledges the support from a Carl-Zeiss Foundation Doctoral Scholarship. We thank Ilja Gerhardt for providing the sodium vapor cells and the oven design. We are grateful to the European Southern Observatory for the loan of a dye-laser system.